# A theory of turbulence based on scale relativity

**Louis de Montera**

**Abstract**

The internal interactions of fluids occur at all scales therefore the resulting force fields have no reason to be smooth and differentiable. The release of the differentiability hypothesis has important mathematical consequences, like scale dependence and the use of a higher algebra. The law of mechanics transfers directly these properties to the velocity of fluid particles whose trajectories in velocity space become fractal and non-deterministic. The principle of relativity is used to find the form of the equation governing velocity in scale space. The solution of this equation contains a fractal and a non-fractal term. The fractal part is shown to be equivalent to the Lagrangian version of the Kolmogorov law of fully-developed and isotropic turbulence. It is therefore associated with turbulence, whereas the non-fractal deterministic term is associated with a laminar behavior. These terms are found to be balanced when the typical velocity reaches a level at which the Reynolds number is equal to one, in agreement with the empirical observations. The rate of energy dissipated by turbulence in a flow passing an obstacle that was only known from experiments can be derived theoretically from the equation's solution. Finally, a quantum-like equation in velocity space is proposed in order to find the probability of having a given velocity at a given location. It may eventually explain the presence of large scale coherent structures in geophysical turbulent flows, like jet-streams.

## 1   Introduction

The classical approaches to tackle the turbulence problem are usually of two kinds. The first one is to consider the Navier-Stokes equations and look at the effect of small perturbations or at the balance between the various terms. The second one is to use dimensional analysis to deduce statistical scale laws. These two methods provide some basic features of turbulence, but they cannot really be considered to be satisfactory since there is no proof that turbulence is indeed a possible solution of the proposed fundamental equations.

The idea presented in this paper is to use the approach of the scale relativity theory and to apply it to fluid mechanics in order to obtain a more general theory that includes turbulence. Compared to classical theories, the foundation of scale relativity is to abandon the hypothesis of differentiability, or smoothness. In the framework of fundamental physics, it is the differentiability of space-time that is seen as an unnecessary and restricting hypothesis. Once differentiability is abandoned, it is shown that space-time has a richer structure that is necessarily fractal at small scales, and which gives rise eventually to quantum mechanics and electromagnetism.

There are strong similarities between the scale relativity theory and fluid mechanics. In this theory, because of fractal geometry, space-time becomes as rough a mountainous landscape, to give an image. In this geometry, the possible paths joining two locations and having the same length are infinite in number. Therefore its geodesics are described as a flow with fluid-like equations. As in the case of turbulence, there is a strong scale dependency that is parameterized by simple statistical laws. There is also a transition scale under which the fractal terms become more important than the classical ones, therefore defining two regimes, like the laminar and turbulent ones in fluid mechanics.

The idea to connect turbulence to the scale relativity theory has already been attempted by Dubrulle (1996) concerning the description of intermittency fluxes and by Célérier (2009) who showed that the scale relativity approach combined with the use of quaternions could lead to a spontaneous arising of a rotational velocity field. This important result shows that the use of higher algebra could be an interesting possibility to be further investigated. However, this work was based on the non-differentiability of the trajectories of fluid particles, which yields a scale law on the displacements of the fluid particles that it not consistent with the observations: it is actually the velocity of fluid particles that verifies a statistical scale law, known as the Kolmogorov law.

In this paper, the scale relativity approach starts from the pressure field which is linked to the velocity of fluid particles through the equations of fluid mechanics. The trajectories remain possibly differentiable, but their derivatives, the velocities, are not necessarily differentiable. It is equivalent to applying the scale relativity approach one order of differentiation below its original presentation in the framework of quantum physics. The theoretical base of this new approach is that the internal forces that act on fluid particles have no particular reason to be smooth since interactions take place at all scales in the fluid. In this way, the obtained scale laws are consistent with the Langrangian version of the Kolmogorov law (Kolmogorov 1941, Landau & Lipschitz 1944) and a transition scale between the fractal and non-fractal behaviour is consistent with the Reynolds number.

## 2   The theory of scale relativity

The theory of scale relativity (Nottale 1993, Nottale 2011 and references therein) starts from a space which has absolutely no characteristics. It is in fact a requirement of the principle of relativity, because if space had any characteristic then something would have to be defined in an absolute way, which would lead to a contradiction. Therefore in this framework, space does not have any particular geometrical property. Even differentiability is not assumed: non-differentiability becomes the general case and differentiability only a particular case.

It can be shown that the release of the hypothesis of differentiability implies that the coordinates in such space are scale-dependent and scale-divergent, or fractal. The proof is based on Lebesgue's theorem which states that a continuous curve with finite length is almost everywhere differentiable. As a consequence, a derivative on such curves is also scale-dependent and does not converge when the scale tends towards zero. It is a bit disturbing to talk about the characteristics of the derivative of a non-differentiable curve. However, there is no contradiction provided that we modify the definition of a derivative. In a non-differentiable framework, the derivative can be computed at any scale, but the classical derivative obtained by taking the limit at very small scales does not exist.

The new fractal geometry has to be taken into account when one considers the reference system, which implies that whatever is observed depends on the scale. Therefore the principle of relativity also applies to the state of scale of the reference system. As a result, the physical quantities are scale-dependent and the form of the equations of physics cannot depend on the scale (otherwise the equations of physics would depend on an arbitrary choice, which is not acceptable).

In the following, space-time is still considered to be differentiable because the scales involved in fluid mechanics are much larger than those involved in quantum physics. However, the trajectories will only be differentiable once, which means that the scale relativity approach will be applied at the level of velocity space.

## 3 The pressure field and its gradient

As explained in the introduction, because the internal interactions in a fluid occur at all scales, there is no particular reason to assume that the resulting fields describing these interactions are smooth and differentiable. Therefore these fields are here assumed to be non-differentiable. This assumption is actually not an assumption, but rather corresponds to the release of an assumption. This point is important because it explains why the proposed theory is able to describe a wider range of phenomena compared to classical theories, like basic turbulence features.

Actually, turbulence clearly tends to create non-differentiable fields, which reinforces the non-differentiability hypothesis. If one considers the Lagrangian version of the Corrsin-Obhukov scale law for temperature (Corrsin 1951, Obukhov 1949):

$$(\Delta T)^2 \approx \Delta t \qquad (1)$$

and assumes an incompressible flow for which the law of ideal gas takes the following form:

$$P \approx T \qquad (2),$$

then, by combining these equations, one obtains:

$$\left(\frac{\Delta P}{\Delta t}\right)^2 \approx \frac{1}{\Delta t} \qquad (3).$$

It is obvious that, in the limit of viscosity, the derivative of the pressure field diverges when the time interval decreases:

$$\lim_{dt \to 0} \left|\frac{dP}{dt}\right| = +\infty \qquad (4).$$

In this study, the proposed approach is actually the reverse one. We start by releasing the constraint of differentiability to allow the turbulent behaviour and its scaling laws to emerge. The release of the differentiability hypothesis has important consequences, in particular for the pressure field.

Firstly, it has to be scale-dependent. It is easy to understand intuitively that a non-differentiable object is not smooth and has many details at all scales, which is the most general definition of fractals. The mathematical proof is given in Nottale (2011) in the section II.5.3. As a consequence, one should write explicitly the scale dependency:

$$P(x, y, z, t) \rightarrow P(x, y, z, t, dx, dy, dz, dt) \tag{5}.$$

Since we released the differentiability hypothesis, its gradient does not strictly speaking exist. However, the infinitesimal calculus can be maintained by defining a new derivative that is also explicitly scale-dependent, like for example (with simplified notations):

$$\frac{dP}{dx}(x, dx) = \frac{P(x+dx, dx) - P(x, dx)}{dx} \tag{6}.$$

This derivative does not include a limit and can be calculated at any scale, but this scale has to be clearly specified. It is also not possible to take the limit of this expression when the scale decreases toward zero because it is not defined.

Secondly, the scale-dependent gradient of the pressure fields has two values. The reason is that the right and left derivative have no reason to be equal like in the differentiable case:

$$\frac{P(x+dx, dx) - P(x, dx)}{dx} \neq \frac{P(x, dx) - P(x-dx, dx)}{dx} \tag{7}.$$

We can therefore define a complex gradient:

$$\nabla P = \text{Re}(\nabla P) + i \, \text{Im}(\nabla P) = \frac{\nabla P_+ + \nabla P_-}{2} + i \frac{\nabla P_+ - \nabla P_-}{2} \tag{8}.$$

## 4 Effect on the velocity of fluid particles

In the absence of any external force, the fundamental law of mechanics applied to a fluid particle states that its acceleration is proportional to the force resulting from the internal interactions within the fluid:

$$m \frac{dv}{dt} = f \tag{9}.$$

Obviously, if pressure gradient diverges, so does the fluid particle acceleration. Therefore, if the pressure field is non-differentiable, neither is the velocity of a given fluid particle. It is as if the particle were evolving in a non-differentiable fractal velocity space.

As a consequence, the Lagrangian velocity is scale-dependent:

$$v(t) \rightarrow v(t, dt) \tag{10}$$

Moreover, its derivative, the acceleration of the fluid particle, has the same properties as the pressure gradient, namely, it is a scale-dependent derivative and has two values:

$$A(t) \rightarrow A(t, dt) \tag{11}$$

and a complex quantity may be defined:

$$\tilde{A} = \frac{A_+ + A_-}{2} + i\frac{A_+ - A_-}{2} \tag{12}$$

by using a complex derivative operator:

$$\frac{\tilde{d}}{dt} = \frac{1}{2}\left(\frac{d_+}{dt} + \frac{d_-}{dt}\right) + \frac{i}{2}\left(\frac{d_+}{dt} - \frac{d_-}{dt}\right) \tag{13}$$

Another important consequence is that the fluid particle evolves in a fractal velocity space that can be visualised as a mountainous landscape. If one wants to go from point A to point B across a mountain range, there is an infinity of possible paths of equal distance due to the fractality of the ground, from the highest peaks to the smallest grain of sand. Therefore a fluid description in velocity space should be used here. It means that the acceleration becomes dependent on velocity:

$$\tilde{A}(t, dt) \rightarrow \tilde{A}(v(t, dt), t, dt) \tag{14}$$

## 5   The resultant velocity scale law

Because acceleration is scale-dependent, it evolves with the scale of observation according to a specific scale law. The principle of relativity imposes that this equation cannot be dependent on the scale itself, because it is an arbitrary choice. Therefore, it can depend only on acceleration. The simplest equation in scale space that yields a fractal solution and that verifies the principle of relativity is (in this section the fact that acceleration has two values is omitted for simplicity):

$$\frac{\partial A}{\partial \ln dt} = \alpha + \beta A \tag{15}$$

The multiplicative scale variable is transformed into an additive one by using its logarithm. This point is detailed in Nottale (2011), section II.4.2. The parameters $\alpha$ and $\beta$ do not depend on scale, but may depend on the location. The general solution of this equation is:

$$A = -\frac{\alpha}{\beta} + \gamma dt^{\beta} \tag{16}$$

where $\gamma$ is an integration constant. This expression can be written under the following form:

$$A = a(v,t)\left[1 + \zeta\left(\frac{dt}{\tau}\right)^{1/D-1}\right] \tag{17}$$

where $D$ is the fractal dimension of the trajectory in velocity space and $a$ the non-fractal or classical part of the acceleration. In this expression, $\zeta$ is random variable having a mean equal to zero and whose square has a mean equal to one:

$$\begin{cases} \langle \zeta^2 \rangle = 1 \\ \langle \zeta \rangle = 0 \end{cases} \tag{18}.$$

This random variable arises from the fact that determinism is fundamentally lost because of the fractality of velocity space (see Nottale 2011, section II.5.3). Because there is an infinity of possible paths of equal length in this space, the trajectories have no reason to follow a particular one and therefore they can only be described statistically. In this way, the chaotic and statistical turbulent behaviour emerges from the release of the differentiability hypothesis. The description of the fluid is simpler since it involves fewer assumptions but because the constraints are less, the described phenomena are richer and closer to the observed facts.

The fluctuations of velocity can be obtained simply by multiplying by $dt$:

$$dv = dV + dW = A dt = a\, dt + \zeta\, a \tau^{1-1/D} dt^{1/D} \tag{19}.$$

As it is well known, dimensionality imposes that $D=2$ in the case of the fully-developed isotropic turbulence. Here, this fractal dimension corresponds to pure random walks in velocity space and therefore is the most general (see Nottale 2011). In this case, the fractal part of the velocity fluctuations becomes:

$$dW = \zeta \sqrt{a^2 \tau\, dt} \tag{20}.$$

By taking the square and the mean of this expression, one obtains:

$$\langle dW^2 \rangle = a^2 \tau dt \tag{21}.$$

Then, by setting:

$$\varepsilon = a^2 \tau \tag{22},$$

one recovers the Lagrangian version of the Kolmogorov law in the case of a fully-developed isotropic turbulence where the non-fractal part of velocity vanishes:

$$\langle dv^2 \rangle = \varepsilon dt \tag{23}.$$

The parameter ε can be identified (up to a multiplicative constant) with the density of the energy flux per unit of mass and volume flowing from the large scales to small scales in the energy cascade.

The amount of energy that is necessary for an obstacle to go forward in a fluid can be theoretically deduced. Until now, this relation was only derived from experiments. If one applies dimensional analysis to the definition of ε, one obtains the well known relation between the rate of energy dissipated by turbulence and the characteristics of the obstacle:

$$\varepsilon \approx \left(\frac{U}{\tau}\right)^2 \tau \approx \frac{U^2}{\tau} \tag{24}.$$

Then by introducing the typical size of the obstacle, it yields the expected relation:

$$\varepsilon \approx U^2 \cdot \frac{U}{L} \approx \frac{U^3}{L} \tag{25}.$$

## 6 The transition scale and the onset of turbulence

The transition scale is the scale at which the temporal mean of the fractal and non-fractal parts of the velocity fluctuations have the same weight:

$$\langle dW^2 \rangle = \langle dV^2 \rangle \tag{26}.$$

At the temporal transition scale $dt=\tau$, the temporal mean of the fractal part is given by:

$$\langle dW^2 \rangle = \varepsilon \tau \tag{27}.$$

The non-fractal part is not stochastic, therefore:

$$\langle dV^2 \rangle = dV^2 \qquad (28).$$

Additionally, the rate of dissipation is known to be equal to:

$$\varepsilon = v.|\nabla v|^2 \qquad (29)$$

where $v$ is the kinematic viscosity. Therefore the transition takes place when:

$$v\tau \frac{dV^2}{dx^2} \approx dV^2 \qquad (30).$$

Which yields, by simplifying:

$$\frac{v\tau}{dx^2} \approx 1 \qquad (31).$$

If the typical velocity and length scales are introduced:

$$\begin{cases} dv \approx U \\ dx \approx L \end{cases} \qquad (32)$$

and simply related to the temporal transition scale by:

$$\tau = \frac{L}{U} \qquad (33),$$

the expression becomes:

$$\mathrm{Re} = \frac{UL}{v} \approx 1 \qquad (34).$$

This result is consistent with the experimental observation that, when the Reynolds number is higher than one, the flow is generally turbulent. It is also consistent with the fact that, if the typical length scale is given, for example by the size of an obstacle, the flow switches from laminar to turbulent behaviour when its typical velocity increases.

Here, the fundamental law of mechanics only serves to prove the scale dependence of the velocity fluctuations. Then, the problem can be converted into an equation in scale space which is easily solved. The resultant solution involves a fractal and a non-fractal term, which are found to be balanced when the Reynolds number is equal to one. Thus, the Reynolds number acquires a more general meaning that goes beyond the field of fluid mechanics. It is related to the transition scale that characterises the general scale law followed by scale-dependent physical quantities within the framework of relativity. For example, in the theory of scale relativity itself, this transition scale corresponds to the de Broglie length separating the quantum and classical realms. In the same way, it is a transition between a fractal and a non-fractal domain because it is known that typical paths of quantum mechanical particles are fractal (Feynman & Hibbs 1965). Turbulence can therefore be seen as a quantum realm in velocity space and it may be described by quantum-like equations, but one order of differentiation below the classical ones.

## 7    Possible Schrödinger equation in velocity space

In this section, we will see how the change of geometry in velocity space is responsible for new terms in the equations additionally to the use of a higher algebra.

In velocity space, the derivative of a function $f(v,t)$ takes the following form:

$$\frac{d_\pm f}{dt} = \frac{\partial f}{\partial t} + \frac{\partial f}{\partial v_i} \cdot \frac{dv_{i\pm}}{dt} + \frac{1}{2}\frac{\partial^2 f}{\partial v_i \partial v_j} \cdot \frac{dv_{i\pm} dv_{j\pm}}{dt} \qquad (35).$$

If one takes the mean of this expression, one sees that it does not reduce exactly to its non-fractal part. Because of the scale dependency, the second-order terms cannot be neglected:

$$\frac{\langle dv_{i\pm} dv_{j\pm} \rangle}{dt} = \frac{\langle dW_{i\pm} dW_{j\pm} \rangle}{dt} = \pm\delta_{ij}\varepsilon \qquad (36).$$

Thus, the derivative becomes:

$$\frac{d_\pm f}{dt} = \frac{\partial f}{\partial t} + \frac{\partial f}{\partial v_i}.a_{i\pm} \pm \frac{1}{2}\frac{\partial^2 f}{\partial v_i \partial v_j}.\delta_{ij}\varepsilon \qquad (37).$$

By using the definition of the complex derivative, one obtains a new operator:

$$\frac{\hat{d}f}{dt} = \frac{\partial f}{\partial t} + \tilde{A}.\nabla_v f - i\frac{\varepsilon}{2}\Delta_v f \qquad (38),$$

where the gradient and Laplacian operators function in velocity space:

$$\begin{cases} \nabla_v = \left(\dfrac{\partial}{\partial v_x}; \dfrac{\partial}{\partial v_y}; \dfrac{\partial}{\partial v_z}\right) \\ \Delta_v = \left(\dfrac{\partial^2}{\partial v_x^2} + \dfrac{\partial^2}{\partial v_y^2} + \dfrac{\partial^2}{\partial v_z^2}\right) \end{cases} \qquad (39).$$

To sum up, the non-differentiability of the internal interactions has been transferred to velocity through the law of mechanics. This has led to the description of velocity space as having a fractal geometry. Then this fractal geometry has been taken into account by including new terms in the derivative operator. Now, the law of mechanics may be considered again and the fractal derivative operator can be applied to it in order to obtain a quantum-like equation. However, the non-differentiable part of the internal forces has been included in the geometry of velocity space, so that it no longer appears within the law of mechanics. The effect of viscosity is also neglected at this stage because it only occurs at very small scales, so that the range of scales over which trajectories are non-differentiable is generally wide enough. The new equation writes:

$$m\frac{\hat{d}\tilde{A}}{dt} = \frac{d}{dt}(f_E) \tag{40}$$

where $f_E$ denotes the external forces. If the time derivative of the external forces can be written as a potential in velocity space, one obtains:

$$\frac{\hat{d}\tilde{A}}{dt} = -\frac{\nabla_v \phi}{m} \tag{41}.$$

Then by setting:

$$\tilde{A}(v,t) = -i\varepsilon \nabla_v \left( \ln \psi(v,t) \right) \tag{42}$$

and using the following identity (Nottale 1993):

$$\nabla\left(\frac{\Delta \psi}{\psi}\right) = 2(\nabla \ln \psi . \nabla)(\nabla \ln \psi) + \Delta(\nabla \ln \psi) \tag{43},$$

one obtains a Schrödinger equation in velocity space (for detailed proof, see Nottale 2011, section II.5.6):

$$\frac{\varepsilon^2}{2}\Delta_v \psi + i\varepsilon \frac{\partial \psi}{\partial t} - \frac{\phi}{m}\psi = 0 \tag{44}.$$

The square of the modulus of this equation's solution is interpreted as the probability density in velocity space of the fluid particle velocity. Note that if an external force has a time derivative equal to zero, like gravity, it does not mean that this force has no effect on the quantum-like equation. The related term in the new equation vanishes, but it has to be taken into account at the previous step, when the form of the fractal part of velocity is derived from the law of mechanics. For example, in the case of gravity, the fractal dimension in the vertical should yield an exponent consistent with a Lagragian version of the Bolgiano-Obukhov scale law (Bolgiano 1959, Obukhov 1959).

The Coriolis force may here have an interesting role because its time derivative is not zero. Its time derivative writes:

$$\frac{df_C}{dt} = \frac{d}{dt}\left(-2m\Omega \wedge v\right) = -2m\Omega \wedge a \tag{45}$$

This situation is equivalent to a quantum particle evolving in a magnetic field, but one order of differentiation below. Therefore the presence of a Coriolis effect leads to a nonlinear Schrödinger-like equation in velocity space. The effect of viscosity possesses the same property:

$$\frac{d}{dt}\left(\upsilon \Delta_x v\right) = \upsilon \Delta_x a \tag{46}.$$

If one assumes that the derivative of acceleration in the physical space do not depend on velocity, this term can be transformed easily into a potential in velocity space.

Then one can jump to the Eulerian approach not only in velocity space, but also in physical space. This approach avoids the difficulty of solving directly the Navier-Stokes equations. The law of mechanics is considered in its simple Lagrangian form and the Eulerian approach is used only at the stage of the quantum-like equation, one order of differentiation below. A wave function operating both in velocity and physical space can be defined. In this case, the derivative operator becomes:

$$\frac{d_\pm f}{dt} = \frac{\partial f}{\partial t} + \frac{\partial f}{\partial x_i}.v_i + \frac{\partial f}{\partial v_i}.a_{i\pm} \pm \frac{1}{2}\frac{\partial^2 f}{\partial v_i \partial v_j}.\delta_{ij}\varepsilon \tag{47}$$

and therefore the complex derivative becomes:

$$\frac{\hat{d}f}{dt} = \frac{\partial f}{\partial t} + v.\nabla_x f + \tilde{A}.\nabla_v f - i\frac{\varepsilon}{2}\Delta_v f \tag{48},$$

The new 'convective' term in the quantum-like equation could be also transformed into a potential in velocity space if one assumes that the derivative of acceleration in the physical space do not depend on velocity.

The square of the modulus of the quantum-like equation's solution would then give the probability of having a given velocity in a given location. Thus, this approach provides a framework for solving the problem of finding the mean velocity field in a turbulent flow. It could be used to determine the general circulation in the atmosphere. In the original formulation of the theory of scale relativity in the field of quantum physics, the chaotic aspect of the flow of space-time geodesics gives rise to the Schrödinger equation and therefore to stable 'large scale' structures like atoms. In this framework, chaos and organized structures are not contradictory aspects, but are totally interrelated, because one founds its basement in the other. In the same way, chaotic trajectories of fluid particle in velocity space could be the cause of the arising of large scale stable and localised structures, like jet-streams, with the force of Coriolis playing the role of a magnetic field in velocity space.

## 8  Conclusion

By releasing the hypothesis that interactions between fluid particles are smooth and differentiable, the fluid particles are shown to follow fractal trajectories in velocity space. The scale relativity framework is therefore appropriate to tackle the problem of turbulence. It is shown that the various physical quantities become scale-dependent and then the principle of relativity is used to find the governing equations in scale space. The solutions correspond to velocity fluctuations having a non-fractal and a fractal part associated respectively with the laminar and turbulent behaviour. The fractal part is shown to be consistent with the Lagrangian version of the Kolmogorov law and the transition scale in velocity space is found to occur when the Reynolds number is equal to one. The rate of energy dissipated by an obstacle in a flow is also recovered. Thus, the basic features of turbulence are recovered theoretically in a straightforward way. Finally, a quantum-like equation in velocity space is proposed in which the Coriolis force can play the role of a magnetic field for a quantum particle. Thus the chaotic aspect of turbulence can be linked to the arising of stable structures. This opens the way to further studies explaining how a chaotic process can produce stable localised structures on large time scales, like jet-streams in the atmosphere.

## Acknowledgements

The author wishes to thank Laurent Nottale for helpful comments.## Acknowledgements

The author wishes to thank Laurent Nottale for helpful comments.